\documentclass[prd,preprintnumbers,showpacs,floatfix,nofootinbib,twocolumn]{revtex4}
 \usepackage[dvips,final]{graphicx}
  \usepackage{amssymb}
   \usepackage{amsmath}
    \usepackage{epsfig}
     \usepackage{bm}
      \usepackage{pifont}
\usepackage{color}
 
  \newcommand{\BluTn}[1]{\textcolor{blue}{#1}}
   \newcommand{\RedTn}[1]{\textcolor{red}{#1}}

\begin{document}
\thispagestyle{empty}
\date{February 9, 2006}
\preprint{\hbox{RUB-TPII-18/05}}

\title{Tagging the pion quark structure in QCD\\}
\author{A.~P.~Bakulev}
 \email{bakulev@theor.jinr.ru}
  \affiliation{Bogoliubov Laboratory of Theoretical Physics, JINR,
               141980 Dubna, Russia\\}

\author{S.~V.~Mikhailov}%
 \email{mikhs@theor.jinr.ru}
  \affiliation{Bogoliubov Laboratory of Theoretical Physics, JINR,
               141980 Dubna, Russia\\}

\author{N.~G.~Stefanis\footnote{On leave of absence from
   Institut f\"{u}r Theoretische Physik II,
               Ruhr-Universit\"{a}t Bochum,
               D-44780 Bochum, Germany}}
 \email{stefanis@tp2.ruhr-uni-bochum.de}
  \affiliation{Bogoliubov Laboratory of Theoretical Physics, JINR,
               141980 Dubna, Russia\\}

\begin{abstract}
We combine the constraints on the pion quark structure available
from perturbative QCD, nonperturbative QCD (nonlocal QCD sum rules
and light cone sum rules) with the analysis of current data on
$F_{\pi\gamma\gamma^*}(Q^2)$,
including recent high-precision lattice calculations
of the second moment of the pion's distribution amplitude.
We supplement these constraints with those extracted from the
renormalon approach by means of the twist-four contributions
to the pion distribution amplitude in order to further increase
stability with respect to related theoretical uncertainties.
We show which regions in the space of the first two non-trivial
Gegenbauer coefficients $a_2$ and $a_4$ of all these constraints
overlap, tagging this way the pion structure to the highest degree
possible at present.
\end{abstract}

\maketitle

\begin{figure*}[t]
\vspace*{-2mm}
 \centerline{\includegraphics[width=0.37\textwidth]{%
   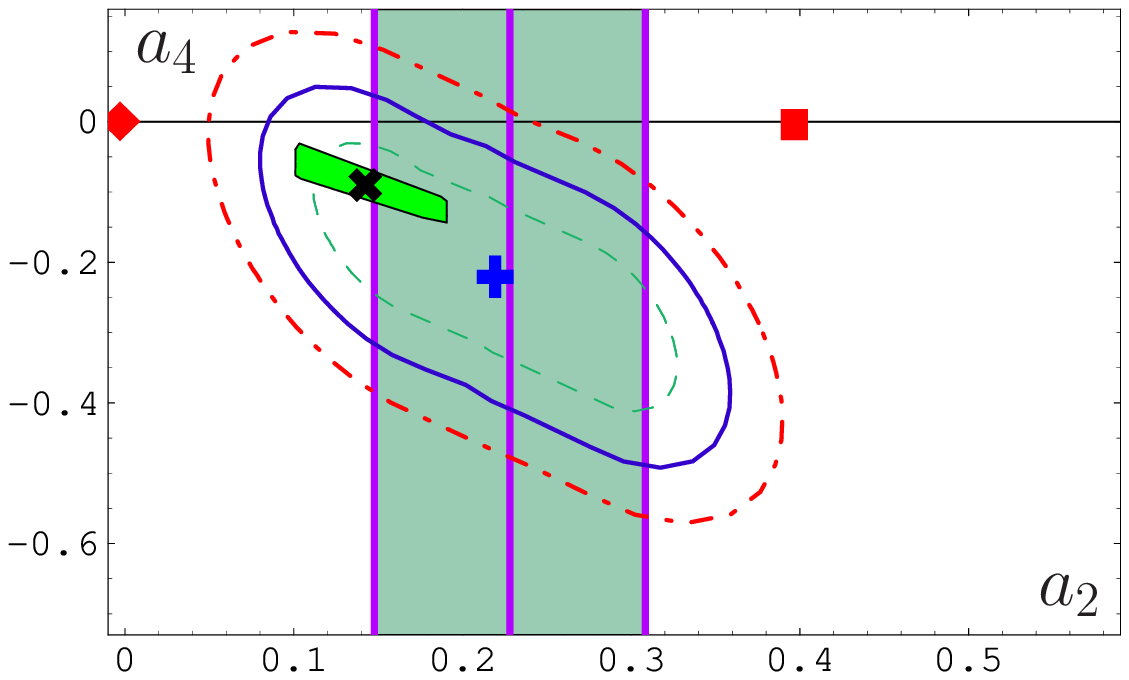}~~~~~~~~%
  \includegraphics[width=0.37\textwidth]{%
   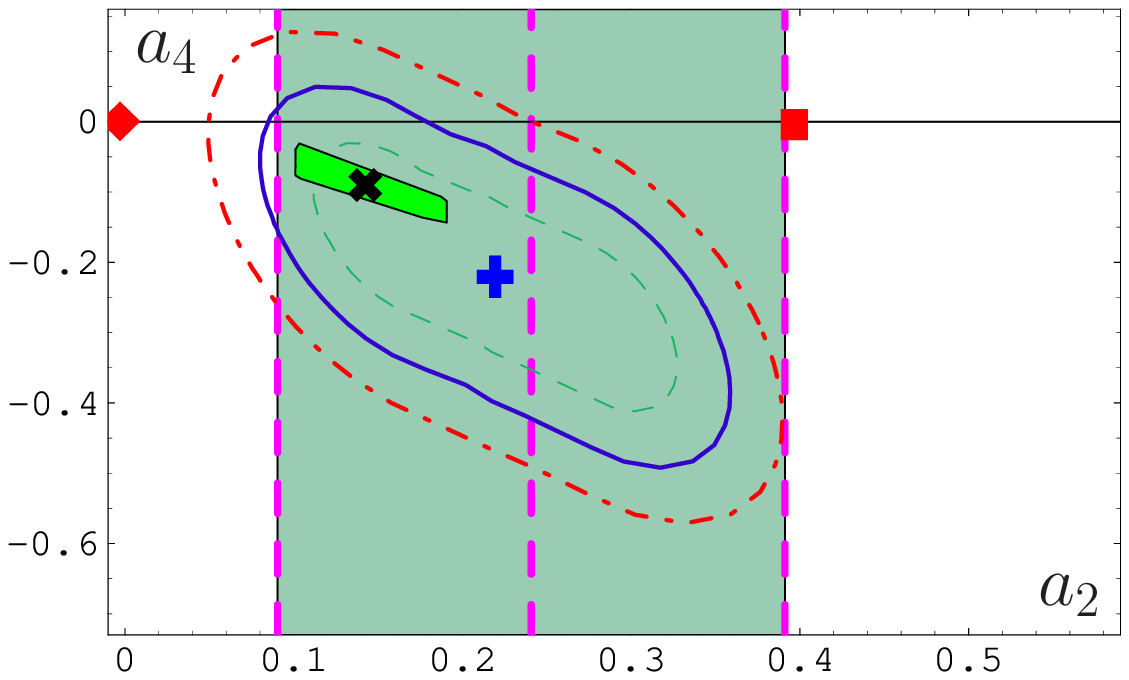}}
   \vspace{0.0cm}
  \centerline{\includegraphics[width=0.37\textwidth]{%
   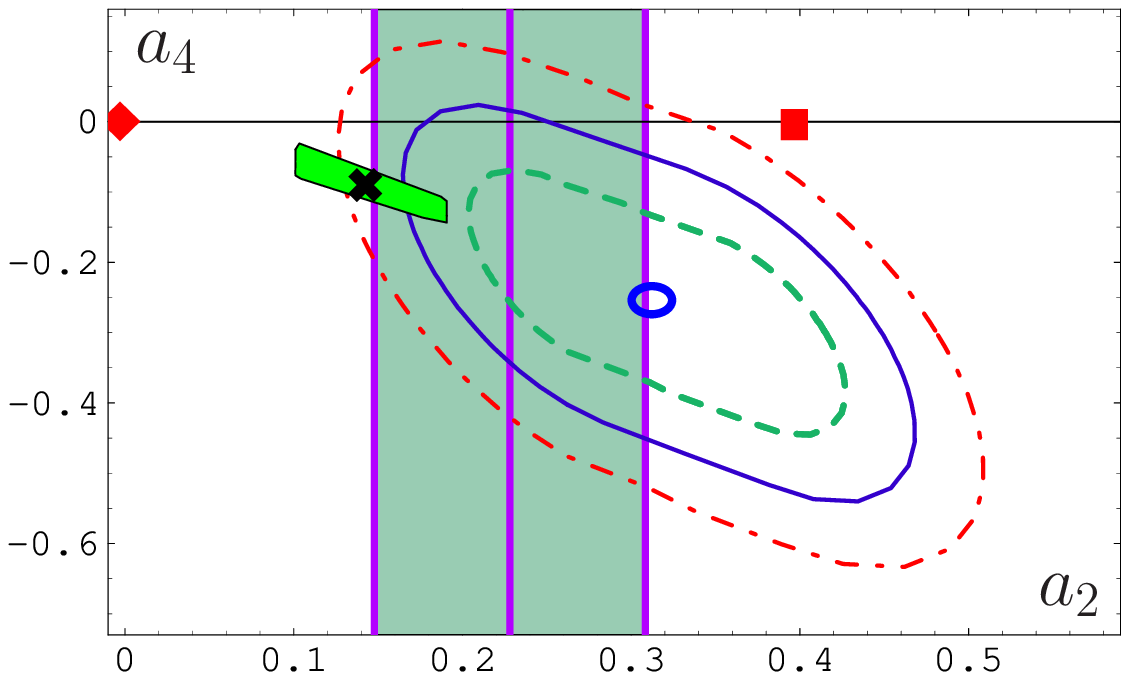}~~~~~~~~%
   \includegraphics[width=0.37\textwidth]{%
   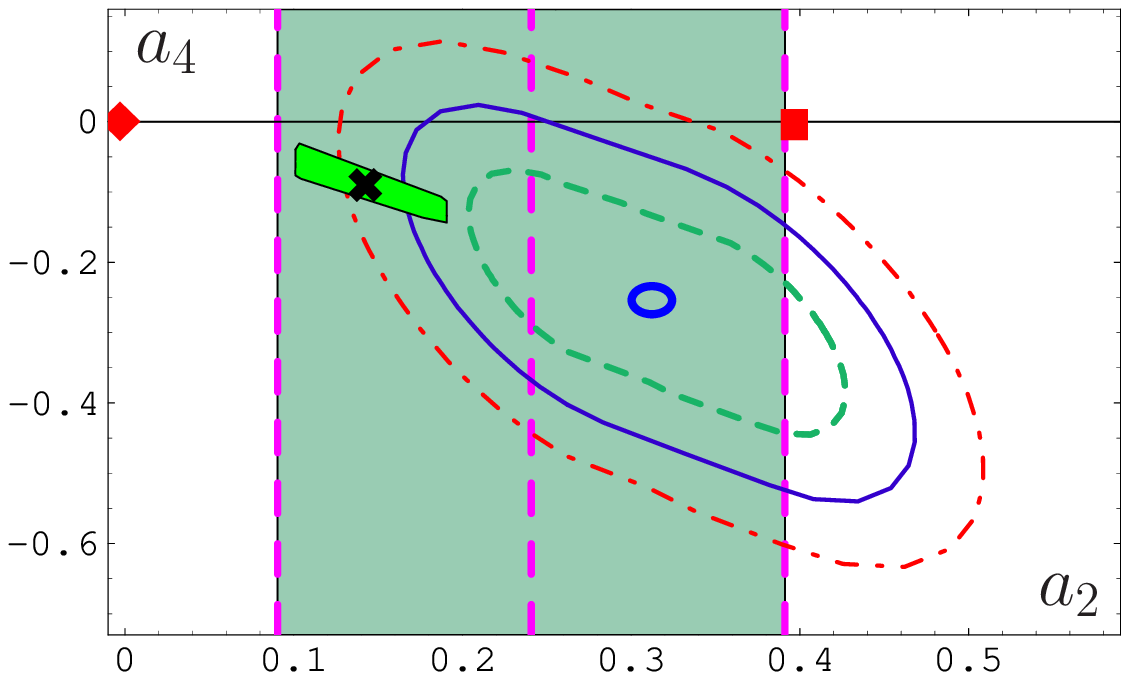}}
   \caption{The top figures show the LCSR-based CLEO-data
   analysis \protect{\cite{BMS02,BMS03}} on
   $F_{\pi\gamma^{*}\gamma}(Q^2)$ in comparison
   with the lattice results of \protect\cite{Lat05} (left
   panel) and \protect\cite{DelD05} (right panel).
   The displayed models are explained in Table
   \protect\ref{tab:collage}.
   The bottom figures show the analogous results for the present
   data analysis which includes twist-four contributions by means of
   the renormalon approach. Here we use: dashed line:
   $1\sigma$; solid line: $2\sigma$;
   dash-dotted line: $3\sigma$); slanted shaded rectangle: confidence
   region of NLC QCD SRs
   \protect\cite{BMS01} for $\lambda^2_q=0.4$~GeV$^{2}$.
    All results are evaluated at
   $\mu^2_\text{SY}=5.76~\text{GeV}^2$
   after NLO ERBL evolution.\label{fig:BMS-lattice}
   }\vspace*{-6mm}
\end{figure*}

The pion distribution amplitude (DA), $\varphi_{\pi}$, \cite{Rad77}
plays a key role in hard-scattering QCD processes because it
encapsulates the essential nonperturbative features of the pion's
internal structure in terms of the partons' longitudinal momentum
fractions $x_i$.
Many studies
\cite{KR96,SY99,SSK99,AriBro-02,BM02,BMS02,BMS03,Ag05a} have
been performed in the literature to determine this DA using the
high-precision CLEO data \cite{CLEO98} on the pion-photon
transition, or to infer it from the calculation of the
corresponding form factor $F_{\pi\gamma^{*}\gamma}(Q^2)$.
In particular, in \cite{BMS03} we have used light cone sum rules
(LCSR) \cite{Kho99,SY99} with a spectral density
obtained in the standard factorization scheme
at the next-to-leading-order (NLO) of perturbative QCD.
We examined the theoretical uncertainties involved in the CLEO-data
analysis in order to extract more reliably the first two non-trivial
Gegenbauer coefficients $a_2$ and $a_4$, which parameterize the
deviation from the asymptotic expression
$\varphi_{\pi}^\text{asy}=6x\bar{x}$
(where $\bar{x}\equiv 1-x$).
A major outcome of this study \cite{BMS03} was that the best agreement
with the data is provided by a pion DA with
$a_2^\text{best fit}(\mu_\text{SY}^2)=+0.23$ and $a_4^\text{best
fit}(\mu_\text{SY}^2)=-0.22$ at the scale
$\mu_\text{SY}^2=5.76$~GeV$^2$
(Schmedding and Yakovlev \cite{SY99}), where the twist-four
contribution was represented by its asymptotic form.
This best-fit solution (\BluTn{\ding{58}}) turns out to be very close
to the model we derived before in \cite{BMS01}---dubbed BMS model
({\footnotesize\ding{54}})---from QCD sum rules with nonlocal
condensates (NLC QCD SR)---see Fig.\ \ref{fig:BMS-lattice}.
Note that this model DA yields a prediction for
$F_{\pi\gamma^{*}\gamma}(Q^2)$ that is in good agreement
\cite{BMS03efr,BMS04kg} not only with the CLEO data, but also with
the older low-$Q^2$ CELLO~\cite{CELLO91} data.
The profile of this pion DA is double-humped, but has its endpoints
$x=0$ and $x=1$ strongly suppressed.
This suppression is directly related to the vacuum quark virtuality
$\lambda_{q}^2$, whose value $0.4$~GeV$^2$, extracted from the CLEO
data in \cite{BMS02}, coincides with that used in \cite{BMS01}, the
reason being that, in the latter, it guarantees best stability of
the NLC QCD SRs employed.
As one sees from Fig.\ \ref{fig:BMS-lattice}, the region of the
$a_2$ and $a_4$ values, admitted by the NLC QCD SRs (small
slanted rectangle), is almost entirely enclosed by the
$1\sigma$-error ellipse. The exclusion of the Chernyak--Zhitnitsky
(CZ)~\cite{CZ84} pion DA (\RedTn{\footnotesize\ding{110}}) and the
asymptotic one (\RedTn{\ding{117}}), found before by
Schmedding and Yakovlev \cite{SY99} at the $2\sigma$ level, was
reinforced in \cite{BMS02}.
In the updated analysis \cite{BMS03}, which accounts for the variation
of the size of the twist-four uncertainties more precisely, these pion
DAs were found to be outside the $4\sigma$ and $3\sigma$ ellipses,
respectively.

Recently, high-precision lattice measurements of the second moment
$\langle{\xi^2\rangle}_{\pi} = \int_0^1(2x-1)^2\varphi_\pi(x)\,dx$
of the pion DA were reported by two different collaborations
\cite{DelD05,Lat05}.
Both groups extracted from their respective simulations, values of
$a_2$ at the scale $\mu^2_\text{SY}$, but with different error
bars.\footnote{We have evolved at the two-loop level both lattice
results to the scale $\mu^2_\text{SY}$ in order to facilitate
comparison with the CLEO-data analyses.}
Unfortunately, the determination of the value of $a_4$ via the moment
$\langle{\xi^4}\rangle_{\pi}$ appears to be a very difficult task on
the lattice because it involves four covariant derivatives
\cite{Lat05}.
It is remarkable that these lattice results are in striking agreement
with the previous estimates of $a_2$ both from NLC QCD SRs
\cite{BMS01} and also from the CLEO-data analyses---based on
LCSR---\cite{SY99,BMS02,BMS03}, as illustrated in Fig.\
\ref{fig:BMS-lattice}, where the lattice results are shown in the form
of a vertical strip, containing the central value with associated
errors.
The (preliminary) results of the QCDSF/UKQCD collaboration \cite{Lat05}
are shown in the left panels of Fig.\ \ref{fig:BMS-lattice}.
The right panels of this figure show the latest estimates by Del Debbio
et al., reported in \cite{DelD05}.
Previous determinations of the same team \cite{DelD02} favored the
asymptotic DA, on the basis of much smaller values of the second moment
$\langle{\xi^2}\rangle_{\pi}$.
It seems that the higher the precision of their simulation, the larger
the result for $\langle{\xi^2}\rangle_{\pi}$ and, consequently, the
more the asymptotic pion DA is disfavored.
Remarkably, the value of $a_2$ of both displayed lattice measurements
(middle line of each strip) is very close to the CLEO best fit in
\cite{BMS02,BMS03} (\BluTn{\ding{58}}), whereas the model pion DA
(\ding{54}) \cite{BMS01} is either inside the error margins (Del
Debbio et al.), or just on the boundary (QCDSF/UKQCD collaboration).

But the success of the CLEO-data analysis depends crucially on the
progress in reducing theoretical uncertainties originating from
uncalculated (perturbative) higher-order radiative corrections and
(mainly nonperturbative) higher-twist contributions.
Those next-to-next-to-leading (NNLO) corrections proportional to the
lowest $\beta$-function coefficient in the standard scheme were
recently calculated in \cite{MMP02} and addressed in \cite{BMS03}.
The higher-twist contributions encode information on power-suppressed
corrections, related to higher quark-gluon Fock states and the quark
transverse momentum.
Just recently, Agaev \cite{Ag05b} has attempted to improve the
CLEO-data analysis by including twist-four contributions, specified by
the renormalon approach~\cite{And99,Ag04,BGG04}.
The main aim was to estimate an upper bound for this contribution and
extract new constraints on the Gegenbauer coefficients $a_2$ and $a_4$
of the leading twist-two pion DA.
Agaev claims that the renormalon approach excludes negative values of
$a_4$.
This finding, remarkable if true, is incompatible with the results
derived from the analyses in \cite{SY99,BMS01,BMS02,BMS03}, discussed
above, and, therefore, deserves closer scrutiny.

Retracing Agaev's calculation, we found that it is seriously flawed in
several respects.
Firstly, we revealed that the normalization of the twist-four pion DAs
is wrong because it erroneously depends on the Gegenbauer coefficient
$a_4$ instead of being a constant.
This might be the root cause for excluding negative values of $a_4$.
Secondly, the Gegenbauer coefficients evolved to the scale
$\mu_\text{SY}^2$, given in his equation (4.2), turn out to be
inconsistently low.
These errors aside, this sort of approach looks quite attractive to be
followed and rectified.
In an effort to carry the renormalon approach to its logical conclusion
and compare its predictions with the previous ones, we take recourse in
the present investigation to the robust results of
Braun--Gardi--Gottwald~\cite{BGG04}, which generalize the previous work
of Andersen in \cite{And99}, and embed their twist-four contributions
to the pion DA into our data-processing framework \cite{BMS02,BMS03}.

The starting point of our discussion is the twist expansion of
the matrix element that defines the pion DAs (e.g., \cite{BF89}):
\begin{widetext}
\begin{eqnarray}
\label{eq:PiDA-Tw4}
  \langle 0 \vert \Big[\bar{d}(z) \gamma_{\nu}\gamma_{5}
         {\cal W}(z,0) u(0)\Big]_{\mu_\text{F}^2}\vert \pi^+(p)
  \rangle
&=& i\,f_\pi \!\!\int_0^1\!\!dx\, e^{-i\bar{x}pz}
          \Big\{p_\nu
              \left[\varphi^{(2)}(x;\mu_\text{F}^2)
              + z^2\varphi^{(4)}_1(x;\mu_\text{F}^2)\right]
\nonumber\\
&& \hspace*{-3pt}
    +\, \Big(z_\nu(pz)-p_{\nu}z^2\Big)\varphi^{(4)}_2(x;\mu_\text{F}^2)
          \Big\} + {\cal O}(z^4)\, ,
\end{eqnarray}
\end{widetext}
where the scale $\mu_\text{F}^2$ defines the factorization scale.
Gauge invariance is ensured by the insertion of the light-like Wilson
line
${\cal W}(z,0)$
$=$
$ P \exp\left[-ig\int_0^1\!dt\,z_\mu A^\mu(z -t z)\right]$.
The twist-two contribution to the pion DA can be represented as an
expansion in terms of Gegenbauer polynomials to read
\begin{equation}
\label{eq:PiDA_2-parameters}
 \varphi^{(2)}(x)
  = \sum_{n=0}a_n \psi_n(x)\,;~
    \psi_n(x) =  6x \bar{x} \cdot C^{3/2}_n(2x-1)\,.
  \end{equation}
For the leading twist-two part, we use a two-parameter model with the
coefficients $a_2$ and $a_4$.
This model receives support from nonlocal QCD sum rules \cite{BMS01}
(cf.\ Fig.\ \ref{fig:BMS-lattice}).
At the reference scale $\mu^2_\text{SY}$, we have after NLO evolution
$a_2^\text{BMS}=+0.14$, $a_4^\text{BMS}=-0.09$.
Note that though the NLO evolution generates higher Gegenbauer
harmonics in the pion DA, the corresponding coefficients remain
numerically negligible, so that the analysis can be based only on
$a_2$ and $a_4$ \cite{BMS02}.
To estimate the twist-four contribution, we now use the renormalon
approach of Ref. \cite{BGG04}.
To this end, we express the corresponding DA, $\Phi^{(4)}(x)$,
characteristic for the reaction $\gamma^* \gamma \to \pi^0$
\cite{Kho99}, with the aid of $\varphi^{(2)}(x)$, in the form of a
convolution
\begin{eqnarray}
 \label{eq:convolution}
  \Phi^{(4)}(x;\mu^2)
   &=& \delta^2(\mu^2)\,K(x,y) \otimes \varphi^{(2)}(y)
   \nonumber\\
   &\equiv& \delta^2(\mu^2)
        \int_0^1\!\! K(x,y)\,\varphi^{(2)}(y)\,dy
  \,,
 \end{eqnarray}
where the kernel $K$ is determined in the present work to be
\begin{eqnarray}
\label{eq:kernel}
  K(x,y)
  &=&
  -\frac{2}{3}
    \Big\{\theta(y>x)\left[\frac{x \bar{x}}{y^2}
  + \frac{1}{y}\ln\left(1-\frac{x}y \right)\right]
  \nonumber\\
  && ~~~~~~+ (x \to \bar{x}, y \to \bar{y})
              \Big\}\,.~~~~
\end{eqnarray}
In Eq.\ \ref{eq:convolution}, the coupling $\delta^2(\mu^2)$ is
defined by (see, for instance, \cite{Kho99})
$\langle \pi (p)|g_\text{s} \bar{d}\tilde{G}_{\alpha\mu}
  \gamma^{\alpha} u|0\rangle=i \delta^2f_{\pi} p_\mu$,
where
$
 \tilde{G}_{\alpha\mu}
=
(1/2)\varepsilon_{\alpha\mu\rho\sigma}G^{\rho\sigma}
$
and $G_{\rho\sigma}=G_{\rho\sigma}^a \lambda^a/2$.
The normalization equation for the twist-four contribution, following
from Eqs.\ (\ref{eq:convolution}) and (\ref{eq:kernel}), is
\begin{eqnarray}
 \label{eq:normt4}
  \int_0^1\!\! d\,x\, K(x,y)\otimes \psi_n(y)
   = \frac{8}9\,\delta_{0n}\,.
\end{eqnarray}
It is important to notice in this context that Eq.\
(\ref{eq:convolution}) accumulates the content of the twist-four pion
DAs, termed
$\varphi_{1}^{(4)},~~\varphi_{2}^{(4)},~~\Phi_{\|},~~\Psi_{\|}$
in \cite{Ag05b} (Eq.\ (3.4) there)---see also Eq.\ (20) in
\cite{Kho99}.
Explicit expressions for these DAs can be found in \cite{BGG04}
(Eqs.\ (3.4) and (3.5)).
Here we display the explicit expression for the twist-four DA we
have derived:
\begin{widetext}
\begin{eqnarray}
 \label{eq:Tw-4_DA}
  \Phi^{(4)}(x;\mu^2)
   &=& -4\,\delta^2(\mu^2)\,
        \left\{x^2\ln(x)(1+a_2+a_4)
               - 5 a_2\,x\bar{x}\,
             \left[\frac{1}{2} - \frac{1}{4}x\bar{x} + 2 x\ln(x)\right]
        \right.
\nonumber\\
   && ~~~~~~~~~~~~~~~\left.
               -\ 7 a_4\,x\bar{x}\,
                \left[1
                    + \frac{35}{8}x\bar{x}
                    - \frac{47}{4} (x\bar{x})^2
                    + \left(13-27x+18 x^2\right) x\ln(x)
                \right]
               + \left(x\to \bar{x}\right)
          \right\}\,.
\end{eqnarray}
\end{widetext}

\begin{figure}[b]
 \centerline{\includegraphics[width=0.4\textwidth]{%
   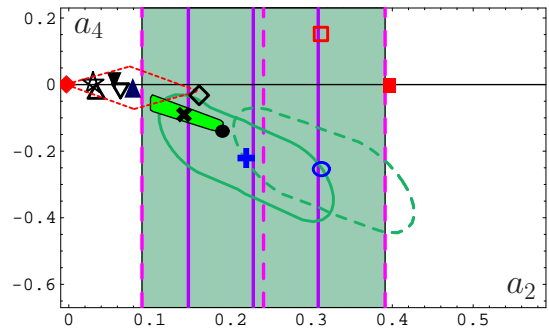}}
   \vspace{0.0cm} \caption{Collage of several proposed models, compiled
   in Table \protect\ref{tab:collage}, in comparison with the CLEO data
   and lattice measurements.
   The dashed rhombus containing the BZ point represents the BZ
   constraints \protect\cite{BZ05}.
   All results are evaluated at
   $\mu^2_\text{SY}=5.76~\text{GeV}^2$
   after NLO ERBL evolution.
\label{fig:final}}
\end{figure}

We estimate within this renormalon-based approach error ellipses of the
CLEO data and show the results of the data processing in
Figs.\ \ref{fig:BMS-lattice} and \ref{fig:final}.
As one sees from these figures, the general effect, entailed by the
inclusion of the twist-four contributions by means of the renormalon
approach, is to shift the error ellipses further to the right,
somewhat enlarging them as well.
One observes that larger values of $a_2$ and $a_4$ are preferred,
while, at the considered $1\sigma$ error level, the sign of $a_4$
remains unchanged and negative.
This phenomenon is in agreement with what we had found before in
\cite{BMS03}, see Table 1 there, when we varied the twist-four coupling
parameter $\delta(\mu^2)$ from smaller to larger values.

\begin{table*}[t]
\caption{Estimates at the normalization scale $\mu_0^2=1$~GeV$^2$ for
 the Gegenbauer coefficients and the reduced inverse moment
 $\langle x^{-1} \rangle^\text{R}_{\pi}$
 for several model DAs for the pion, based on the first two Gegenbauer
 coefficients $a_2$ and $a_4$. The designations correspond to those
 used in Fig.\ \protect\ref{fig:final}.
 Included are also the theoretical constraints derived from LCSRs and
 NLC QCD SRs, and by analyzing~\cite{BMS02} the CLEO data
 \cite{CLEO98}.
  Also shown are the lattice measurements of \protect\cite{DelD05} and
 \protect\cite{Lat05} and the previous transverse-lattice result.
 [Note that the uncertainties on the Gegenbauer coefficients $a_2$ and
 $a_4$ are correlated.
 Here, the rectangular limits of the fiducial ellipse
 \protect\cite{BMS02,BMS03} are shown.]
 \label{tab:collage}}
\begin{ruledtabular}
\begin{tabular}{ccccc}
DA models/methods
 & Symbols
    & $a_2(\mu_0^2)$
        & $a_4(\mu_0^2)$
           & $\langle x^{-1} \rangle^\text{R}_{\pi}$
             \\ \hline \hline
As
 & \RedTn{\ding{117}}
    & 0
        & 0
           & 0
              \\
ADT~\cite{ADT00}
 & {\footnotesize$\bigtriangleup$}
    & $0.05$
        & $-0.04$
           & $0.01$
              \\
BMS best fit point~\cite{BMS02}
 & \BluTn{\ding{58}}
    & $0.31$
        & $-0.35$
           & $-0.04$\phantom{$-$}
              \\
Best fit point (ren)
 & \BluTn{\Large$\circ$}
    & $0.44$
        & $-0.40$
           & $0.04$
              \\
PPRWG~\cite{PPRWG99}
 & {\ding{73}}
    & $0.046$
        & $0.007$
           & $0.05$
               \\
SY~\cite{BF89}
 & {\footnotesize\ding{108}}
    & $0.27$
        & $-0.22$
           & $0.05$
              \\
BMS~\cite{BMS01}
 & {\ding{54}}
    & $0.20$
        & $-0.14$
           & $0.06$
              \\
PR~\cite{PR01}
 & {\footnotesize$\bigtriangledown$}
    & $0.09$
        & $-0.02$
           & $0.07$
              \\
BZ~\cite{BZ05}
 & {\footnotesize\ding{115}}
    & $0.12$
        & $-0.02$
           & $0.10$
              \\
Agaev~\cite{Ag05a}
 & {\Large$\diamond$}
    & $0.23$
        & $-0.05$
           & $0.18$
               \\
CZ~\cite{CZ84}
 & \RedTn{\footnotesize\ding{110}}
    & $0.56$
        & 0
           & $0.56$
               \\
BF~\cite{BF89}
 & \RedTn{$\Box$}
    & $0.44$
        & 0.25
           & $0.69$
               \\ \hline\hline
NLC QCD SRs for $\langle{\xi^N}\rangle_\pi$
\protect{\cite{BMS01,BMS03}}
 & \begin{minipage}[c]{60pt}
    $\includegraphics[width=30pt]{
    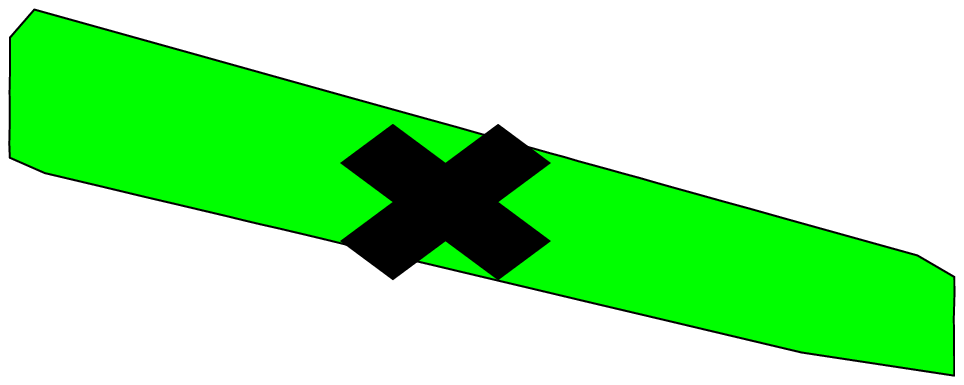}$
    \end{minipage}
     & $[0.13, 0.25]$
        & $[-0.04, -0.22]$
           & ---
               \\
``Daughter'' NLC QCD SR \protect{\cite{BMS01,BMS03}}
 &
    & ---
        & ---
           & $0.09 \pm 0.10$
               \\
LCSR analysis of JLab data \protect{\cite{BiKho02}}
 &
    & $0.24\pm 0.14\pm 0.08$
        & ---
           & ---
               \\
LCSR analysis of CLEO data~\cite{CLEO98,BMS03}
 &
    & ---
        & ---
           & $-0.03 \pm 0.18$\phantom{$-$}
               \\
Transverse lattice~\cite{Dal02}
 & {\footnotesize\ding{116}}
    & $0.08$
        & $0.02$
           & $0.10$
               \\
QCDSF/UKQCD lattice~\cite{Lat05}
 &  \begin{minipage}[c]{60pt}
    $\includegraphics[width=20pt,height=10pt]{
    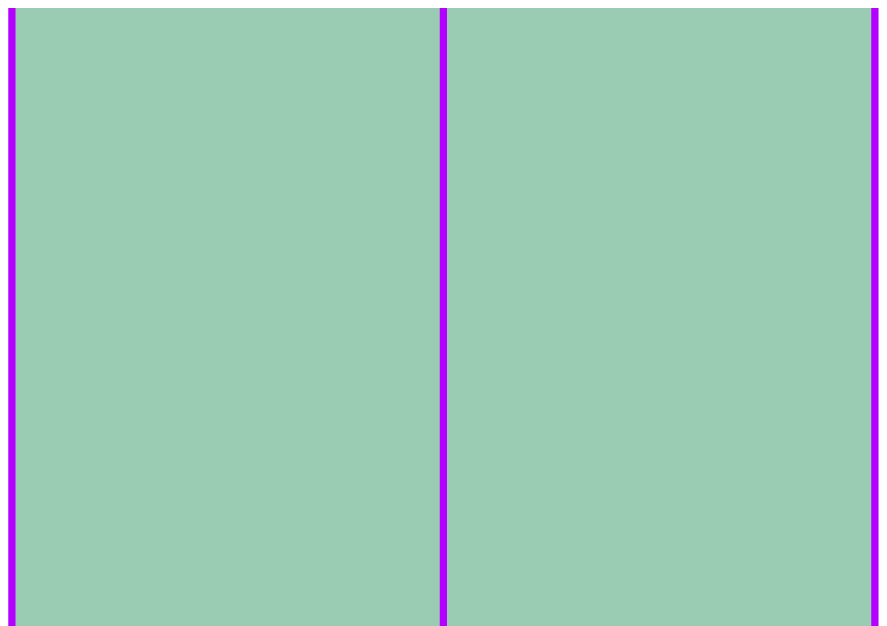}$
    \end{minipage}
    & $0.32\pm0.11$
        & ---
           & ``$0.32\pm0.11$''
               \\
Del Debbio et al. (lattice)~\cite{DelD05}
 &   \begin{minipage}[c]{60pt}
    $\includegraphics[width=30pt,height=10pt]{
    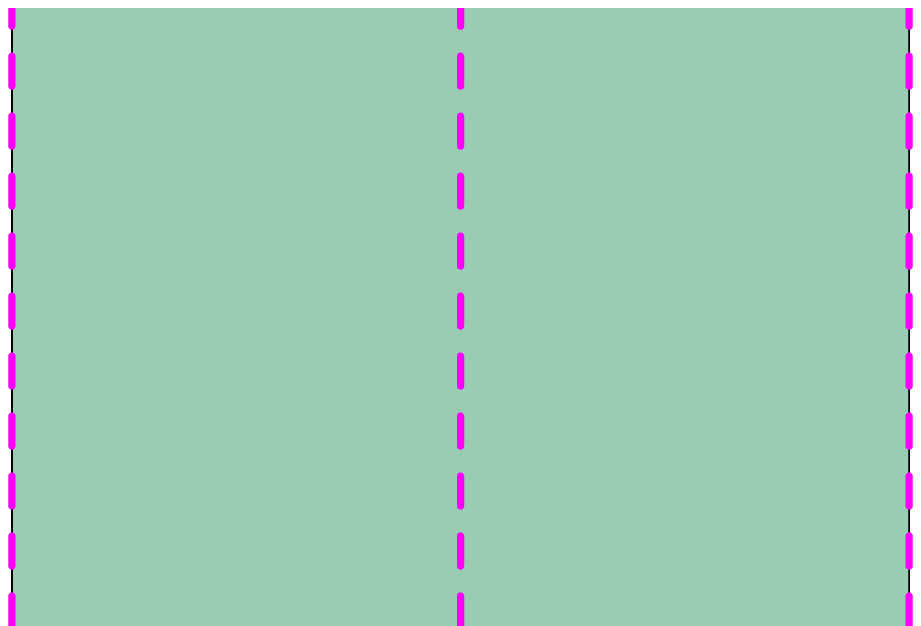}$
    \end{minipage}
    & $0.34\pm0.21$
        & ---
           & ``$0.34\pm0.21$''
\end{tabular}
\end{ruledtabular}
\end{table*}

For the new best-fit point (\BluTn{\Large$\circ$}), we obtain the
following values
$a_2^\text{best fit(ren)}(\mu_\text{SY}^2)=+0.31$
and
$a_4^\text{best fit(ren)}(\mu_\text{SY}^2)=-0.25$.
Note that the $\chi^2$ value of that best-fit point
($\chi^2_\text{best fit(ren)}=0.58$)
is comparable with the one in the standard analysis \cite{BMS02,BMS03}
($\chi^2_\text{best fit}=0.47$).
Analyzing the structure of the twist-four pion DA, given by Eq.\
(\ref{eq:Tw-4_DA}), one finds that it has two distinctive peaks and
strongly enhanced endpoint regions $x=0,1$, in accordance to the
assertions in \cite{BGG04} and \cite{Ag05b}.
But, perhaps somewhat surprisingly, including this contribution into
the LCSR analysis of the CLEO data, the characteristics of the best-fit
pion DA in the endpoint region remain almost the same as found before
\cite{BMS02,BMS03} with the use of the asymptotic form for
$\Phi^{(4)}(x;\mu^2)$.
Both DAs show endpoint suppression, much like the BMS model DA
extracted from the NLC QCD SRs \cite{BMS01}, though these two
approaches are not related to each other.
Given that the renormalon-based twist-four contribution represents an
\emph{upper} bound \cite{BGG04}, we can conclude that the endpoint
region of the pion DA will not receive further enhancement.

Figure \ref{fig:final} presents a collage of several proposed models
for the pion DA, listed in Table \ref{tab:collage}, in comparison with
the CLEO-data constraints in terms of the $1\sigma$ ellipses, and the
recent lattice measurements of Refs.\ \cite{DelD05,Lat05}.
The influence of NNLO radiative corrections---only
partially known \cite{MMP02}---can be investigated
by means of varying the factorization scale,
say, in the interval $Q^2/2\leq\mu^2\leq 2Q^2$,
where $\mu^2=\mu_\text{F}^2=\mu_\text{R}^2$.
This has been discussed in \cite{BMS03} and it has been shown that
it does not alter qualitatively the main results.
Note that the NNLO corrections within the conformal scheme have been
quantitatively analyzed in \cite{MMP02}.
Looking more carefully on Fig.\ \ref{fig:final}, one recognizes an
interesting pattern: with the exception of the BF model and the CZ one,
all other models steer themselves along an ``orbit'' defined by
$a_2+a_4\approx\text{const}$.
The organizing tool behind that pattern is the reduced inverse moment
$\langle x^{-1} \rangle^\text{R}_{\pi}$ of the pion DA,
\begin{equation}
  \langle x^{-1} \rangle^\text{R}_{\pi}
   \equiv
         \frac{\langle x^{-1} \rangle_{\pi}}{3}
    - 1\
    =\ a_2 + a_4 + \ldots\,,
\label{eq:Delta}
\end{equation}
where the ellipsis stands for still higher Gegenbauer coefficients.
This inverse moment plays a special role in the description of several
form factors of the pion in perturbative QCD.

The rationale for it is as follows.
Though the diversity in shape among the models shown in the first
column of Table \ref{tab:collage}---in correspondence with Fig.\
\ref{fig:final}---is substantial, the associated values for the
reduced inverse moment are relatively close to each other (see the
last column of Table \ref{tab:collage}).
Nevertheless, the predictions for the pion-photon transition form
factor vary quite strongly along this orbit as it is evident from
Fig.\ \ref{fig:final}.
The deep reason for this, is that this form factor is \textit{not}
directly proportional to the reduced inverse moment (for more details,
the interested reader is referred to Appendix E of \cite{BMS02}
and to \cite{BMS01}).

In addition, as it can be shown by using the techniques developed in
\cite{BPSS04,SBMPS04}, if the reduced inverse moment is in the range
of the NLC QCD SRs (see Table \ref{tab:collage}), then also the hard
contribution to the electromagnetic pion form factor has the right
magnitude---provided one uses for \emph{all} model DAs, the
\emph{same} soft contribution, say, that employed in
\cite{BPSS04,SBMPS04} on account of quark-hadron local duality.
Moreover, the estimate
$\displaystyle \langle x^{-1} \rangle^\text{EM}_{\pi}/3-1=0.24\pm 0.16$,
obtained in the data analysis of the electromagnetic (EM) pion form
factor within the framework of a different LCSR method in
\cite{BKM00,BiKho02}---though only dependent on $a_2$---is compatible
within errors with the quoted NLC QCD SR result.
On that basis, one might expect that future lattice measurements should
yield \emph{negative} values of $a_4$ in order to be in line with the
typical value of the reduced moment.
This would decrease the values shown in Table \ref{tab:collage} in
quotation marks, recalling that they contain only the first
Gegenbauer coefficient $a_2$.

We have given a comprehensive compendium of the current situation on
the constraints imposed on the first two non-trivial Gegenbauer
coefficients by the analysis of the CLEO data \cite{CLEO98} in
comparison with the most recent estimates from modern lattice
technology \cite{DelD05,Lat05}.
We used in our data analysis the method of light-cone sum rules
including, on one hand, perturbative corrections up to NLO, and, on the
other hand, taking into account the twist-four contribution to the pion
DA, having recourse to the renormalon approach.
We believe that this investigation further restricts the available
space of values in the $(a_2,a_4)$ plane for the pion DA, confirming
the negative sign of the Gegenbauer coefficient $a_4$ at the
$1\sigma$ level of the CLEO-data analysis without and with the
inclusion of the twist-four contribution via the renormalon method.
The gross profile emerging from this CLEO-data analysis is consistent
with that we have determined independently some years ago \cite{BMS01}
from QCD sum rules with nonlocal condensates.
In fact, all constraints we have presented above are consistent with
a profile for the pion DA that has two humps, but with strongly
suppressed endpoints at $x=0$ and $x=1$.
The endpoint-enhancement of the renormalon-based twist-four
contribution does not change these characteristics.
We emphasize that this type of pion DA leads to a pion-photon
transition form factor that was found \cite{BMS03efr,BMS04kg}
to be in good agreement with the (low) $Q^2$ CELLO data \cite{CELLO91}
as well.
A full-blown reconstruction of the pion DA in terms of its Gegenbauer
decomposition may become possible in the next years, once the lattice
community can provide estimates for $a_4$ that seems quite difficult
at present.
Further advances in the calculation of the higher-twist contributions
to the pion DA and inclusion of the partial NNLO radiative corrections
in the standard scheme \cite{MMP02} will also improve the theoretical
precision in assessing the pion structure via the CLEO or other data
of similar precision.
In this respect, a combined use of data on the pion-photon transition
with such on the electromagnetic pion form factor, after the upgrade to
12~GeV, of the CEBAF machine at JLab, will considerably help to tag
the pion structure at a still deeper level.

Two of us (A.P.B. and S.V.M.) are indebted to Prof.\ Klaus Goeke
for the warm hospitality at Bochum University, where part of this work
was carried out.
N.G.S. thanks for support the BLTP@JINR, where this work was completed.
This investigation was supported in part by the Heisenberg-Landau
Programme (grant 2005) and the Russian Foundation for Fundamental
Research (grant No.\ 03-02-04022).


\begin{thebibliography}{10}
\bibitem{Rad77} A.~V. Radyushkin,
 Dubna preprint P2-10717, 1977 [hep-ph/0410276].

\bibitem{KR96} P. Kroll and M. Raulfs,
 Phys.\ Lett.\ \textbf{B387},  848  (1996).

\bibitem{SY99} A. Schmedding and O. Yakovlev,
 Phys.\ Rev.\ \textbf{D62},  116002  (2000).

\bibitem{SSK99} N.~G. Stefanis, W. Schroers, and H.-C. Kim,
 Phys.\ Lett.\ \textbf{B449},  299  (1999);
 Eur.\ Phys.\ J.\ \textbf{C18},  137 (2000).

\bibitem{AriBro-02} E.~R. Arriola and W. Broniowski,
 Phys.\ Rev.\ \textbf{D66},  094016  (2002).

\bibitem{BM02} A.~P. Bakulev and S.~V. Mikhailov,
 Phys.\ Rev.\ \textbf{D65},  114511  (2002).

\bibitem{BMS02} A.~P. Bakulev, S.~V. Mikhailov, and N.~G. Stefanis,
 Phys. Rev. \textbf{D67}, 074012  (2003).

\bibitem{BMS03} A.~P. Bakulev, S.~V. Mikhailov, and N.~G. Stefanis,
 Phys.\ Lett.\ \textbf{B578},  91  (2004).

\bibitem{Ag05a} S.~S. Agaev,
 Phys.\ Rev.\ D \textbf{72}, 114010 (2005).

\bibitem{CLEO98} J. Gronberg \textit{et~al.},
 Phys.\ Rev.\ \textbf{D57},  33  (1998).

\bibitem{Kho99} A. Khodjamirian,
 Eur.\ Phys.\ J.\ \textbf{C6},  477  (1999).

\bibitem{BMS01} A.~P. Bakulev, S.~V. Mikhailov, and N.~G. Stefanis,
 Phys.\ Lett.\ \textbf{B508}, 279  (2001);
 [Erratum-ibid.\ B \textbf{590}, 309 (2004)];
 in {\em Proceedings of the 36th Rencontres De Moriond on QCD
 and Hadronic Interactions, 17--24 Mar 2001, Les Arcs, France},
 edited by J.~T.~T. Van (World Scientific, Singapore, 2002),
 pp.\ 133--136.

\bibitem{BMS03efr} A.~P. Bakulev, S.~V. Mikhailov, and N.~G. Stefanis,
 Phys.\ Part.\ Nucl.\  \textbf{35},  7  (2004).

\bibitem{BMS04kg} A.~P. Bakulev, S.~V. Mikhailov, and N.~G. Stefanis,
 Annalen Phys.\ \textbf{13}, 629  (2004).

\bibitem{CELLO91} H.~J. Behrend \textit{et~al.},
 Z.\ Phys.\ \textbf{C49},  401  (1991).

\bibitem{CZ84} V.~L. Chernyak and A.~R. Zhitnitsky,
 Phys.\ Rept.\ \textbf{112},  173  (1984).

\bibitem{Lat05} M. G{\"o}ckeler \textit{et~al.},
 hep-lat/0510089.

\bibitem{DelD05} L. Del~Debbio,
 Few Body Syst.\ \textbf{36},  77  (2005).

\bibitem{DelD02} L. Del~Debbio, M. Di~Pierro, and A. Dougall,
 Nucl.\ Phys.\ Proc.\ Suppl.\ \textbf{119},  416  (2003).

\bibitem{MMP02} B. Meli\'{c}, D. M{\"u}ller, and K. Passek-Kumeri\v{c}ki,
 Phys.\ Rev.\ \textbf{D68},  014013  (2003).

\bibitem{Ag05b} S.~S. Agaev,
 hep-ph/0511192.

\bibitem{And99} J.~R. Andersen,
 Phys.\ Lett.\ \textbf{B475},  141  (2000).

\bibitem{Ag04} S.~S. Agaev,
 Phys.\ Rev.\ \textbf{D69},  094010  (2004).

\bibitem{BGG04} V.~M. Braun, E. Gardi, and S. Gottwald,
 Nucl.\ Phys.\ \textbf{B685},  171 (2004).

\bibitem{BF89} V.~M. Braun and I.~E. Filyanov,
 Z.\ Phys.\ \textbf{C44},  157  (1989).

\bibitem{ADT00} I.~V. Anikin, A.~E. Dorokhov, and L. Tomio,
 Phys.\ Part.\ Nucl.\ \textbf{31},  509 (2000).

\bibitem{PPRWG99} V.~Y. Petrov, M.~V.~Polyakov, R. Ruskov, C. Weiss,
 and K. Goeke
 Phys.\ Rev.\ \textbf{D59},  114018  (1999).

\bibitem{PR01} M. Praszalowicz and A. Rostworowski,
 Phys.\ Rev.\ \textbf{D64},  074003  (2001);
 Phys.\ Rev.\ \textbf{D66},  054002  (2002).

\bibitem{BZ05} P. Ball and R. Zwicky,
 Phys.\ Lett.\ \textbf{B625},  225  (2005);
 Phys.\ Rev.\ \textbf{D71},  014015  (2005).

\bibitem{BiKho02} J. Bijnens and A. Khodjamirian,
 Eur.\ Phys.\ J.\ \textbf{C26},  67  (2002).

\bibitem{Dal02} S. Dalley and B. van~de Sande,
 Phys.\ Rev.\ \textbf{D67},  114507  (2003).

\bibitem{BPSS04} A.~P. Bakulev, K. Passek-Kumeri\v{c}ki, W. Schroers,
 and N.~G. Stefanis,
  Phys.\ Rev.\ \textbf{D70},  033014  (2004).

\bibitem{SBMPS04} N.~G. Stefanis \textit{et~al.},
 in {\em First International Workshop ``Hadron Structure and {QCD} (HSQCD 2004):
     From Low to High Energies'', Repino, St.~Petersburg, Russia, 18--22 May 2004},
     edited by V.~T.~Kim and L.~N.~Lipatov
     (PNPI, Gatchina, St.~Petersburg, 2004), pp.\ 238--245 [hep-ph/0409176].

\bibitem{BKM00} V.~M. Braun, A. Khodjamirian, and M. Maul,
 Phys.\ Rev.\ \textbf{D61},  073004  (2000).
\end{thebibliography}

\end{document}